\documentclass[conference]{IEEEtran}
\usepackage{verbatim} 
\usepackage{graphicx}
\usepackage{subfigure}
\usepackage{times}
\usepackage{caption}
\usepackage{epsfig}
\usepackage{stfloats}
\usepackage[noadjust]{cite}
\usepackage{enumerate}
\usepackage{amssymb}
\usepackage{multirow}
\usepackage{latexsym}
\usepackage{color}
\usepackage{amsmath}
\usepackage[noend]{algpseudocode}
\usepackage{epstopdf}
\usepackage{tikz}
\usepackage[american]{circuitikz}
\usepackage{algorithm}
\usepackage{mathtools}

\allowdisplaybreaks
\usepackage{etoolbox}

 \makeatletter 
 \def\@eqnnum{{\normalsize \normalcolor (\theequation)}} 
  \makeatother
\usepackage{etex}
\usepackage{tabularx} 
\usepackage{algorithm}
\usepackage{graphicx}
\usepackage{algpseudocode}
\usepackage{tikz}
\usetikzlibrary{shapes,arrows}
\usepackage{pstricks}
\usepackage{pst-node,pst-blur}
\usetikzlibrary{arrows}
\usepackage{amsfonts}
\usepackage{bm}
\usepackage{tabularx}
\usepackage{enumerate}
\usepackage{subfigure}
\usepackage{amsmath}
\usepackage{amssymb}
\usepackage{booktabs} 
\usepackage{multirow} 
\usepackage{mathtools} 
\usepackage{epsfig}
\usepackage{epstopdf}
\usepackage{array}

\usepackage[noadjust]{cite}
\usepackage{caption}
\usepackage{etoolbox}

\allowdisplaybreaks
\makeatletter
\def\set@curr@file#1{%
  \begingroup
    \escapechar\m@ne
    \xdef\@curr@file{\expandafter\string\csname #1\endcsname}%
  \endgroup
}
\def\quote@name#1{"\quote@@name#1\@gobble""}
\def\quote@@name#1"{#1\quote@@name}
\def\unquote@name#1{\quote@@name#1\@gobble"}
\makeatother
\usepackage{graphics} 

\begin{document}

\title{GITz: Graphene-assisted IRS Design for THz Communication}
\author{\IEEEauthorblockN{Bhupendra Sharma\IEEEauthorrefmark{1}, Anirudh Agarwal\IEEEauthorrefmark{2}, Deepak Mishra\IEEEauthorrefmark{3} and Soumitra Debnath\IEEEauthorrefmark{4}}
\IEEEauthorblockA{\IEEEauthorrefmark{1}\IEEEauthorrefmark{2}\IEEEauthorrefmark{4}Department of ECE, 
The LNM Institute of Information Technology, Jaipur, India\\
\IEEEauthorrefmark{3}School of Electrical Engineering and Telecommunications, University of New South Wales, Sydney, Australia
\\E-mails: bhupendrasharma.y19@lnmiit.ac.in\IEEEauthorrefmark{1}, anirudh.agarwal@lnmiit.ac.in\IEEEauthorrefmark{2}, d.mishra@unsw.edu.au\IEEEauthorrefmark{3}, soumitra@lnmiit.ac.in\IEEEauthorrefmark{4}
}}

% make the title area
\maketitle
% As a general rule, do not put math, special symbols or citations in the abstract or keywords.
\begin{abstract}
 Graphene-based intelligent reflecting surface (GIRS) has been proved to provide a promising propagation environment to enhance the quality of high frequency terahertz (THz) wireless communication. In this paper, we characterize GIRS for THz communication (GITz) using material specific parameters of graphene to tune the reflection of the incident wave at IRS. In particular, we propose a GITz design model considering the incident signal frequency material level parameters like conductivity, Fermi-level, patch width to control the reflection amplitude (RA) at the communication receiver. We have obtained the closed-form expression of RA for an accurate design and characterization of GIRS, which is incomplete in the existing research due to the inclusion of only phase-shift. The numerical simulation results demonstrate the effectiveness of the proposed characterization by providing key insights.   
\end{abstract}

\begin{IEEEkeywords}
Graphene, intelligent reflecting surface, THz communication, reflection amplitude, modeling, characterization. 
\end{IEEEkeywords}

\section{Introduction}
Terahertz (THz) frequency spectrum is the potential candidate to efficiently fulfil the high data rate and network coverage requirement, unlocking high bandwidth for effective spectrum utilization in 6G wireless networks. However, it suffers from propagation loss, signal blockage and significant hardware cost \cite{MIMO_GRAPHENE}. Consequently, it is necessary to explore THz band-related devices with controlled wireless signal propagation. In this context, intelligent reflecting surface (IRS)-assisted THz communication is being universally adopted to address above losses with joint active and passive beamforming design thereby enhancing the quality of service and reducing the communication overheads \cite{sixth_gen_wn}. 

IRS is a promising solution for precise beam orientation towards users at different locations. It contains an array of unit cells; each of which can alter the phase, amplitude and polarization of an incoming signal \cite{Basar}, \cite{sharma2021circuit}. However, the architectural design of an IRS should be optimum in size and mechanically adaptive for easy accessibility. To enhance the performance of IRS in THz and mid-infrared frequency range communication, a special material-specific design is required. 

\subsection{State-of-the-Art}
Metamaterials have gained significant attention due to their periodic arrangement via which the electromagnetic (EM) properties of incident wave can be suitably controlled \cite{emil_b}. In this regard, graphene is emerging as a prominent option for efficient IRS design in THz communication. Its exclusive electrical properties such as high electron mobility and electrical configurability pave the way for high frequency signal transmission. Moreover, graphene has high conductivity which can be tuned by varying the carrier mobility and the Fermi-level \cite{Fermi_level}. Authors in \cite{Hybrid} proposed a hybrid combination of graphene and gold patch at IRS element. They further highlighted the key benefits of adjoining a metal and graphene stub for desired phase response and RA at THz frequency. Authors in \cite{Absorption} found the factors that restrict the absorption and tunability in THz frequency range. More recently, plasmonics of graphene-assisted THz metarials \cite{nature} was studied for enabling wireless communication in THz region. Plasmonic-based antennas assist the propagation of surface plasmon polariton (SPP) waves. The wavelength of SPP waves is much smaller than that of EM waves at same frequency. So, highly dense design of array is possible, which can integrate a massive number of patches in the form of miniature imprint that can be deployed on the IRS structure. The non-line-of-sight links can thus be improved by such a smart array overcoming the complexity of THz communication model. 

Few researchers have also worked on design and modelling of an enhanced graphene-assisted IRS (GIRS) based wireless communication system. Interestingly in \cite{Joint}, the relationship between the conductivity of graphene and applied voltage marked the foundation of developing an electrically-controlled IRS design. Further, authors in \cite{Ysur} presented graphene based artificial magnetic conductor used for the realization of the high impedance surface which is adjusted by tuning the physical parameters of the design. 

\subsection{Research Gap and Motivation}
The research in the field of GIRS-aided THz communication (GITz) is still in its infancy. There is a need to explore mathematical frameworks for quantifying, thereby optimizing the performance of such communication systems. Most of the existing works in GIRS-assisted wireless communication \cite{Joint}, \cite{Ysur} considered phase-shift (PS) by tuning the surface conductivity of graphene for their modeling purpose. \cite{Ysur} investigated  controllable surface conductivity of graphene for dual band high impedance surface in THz region. Similarly in \cite{Joint}, authors proposed joint hardware design and capacity characterization of GIRS-enabled THz multiple-input multiple-output system, thereby adjusting the PS using the material properties of graphene. Apart from PS adjustment, it is pertinent to control RA at the receiving end of GITz system. However, above models neither derived a closed-form expression for RA which is necessary for analyzing the overall performance of a GITz, nor they incorporated the effect of other material-specific parameters for designing the GIRS unit cell, which greatly impacts the quality of high frequency received signal. 

\subsection{Novelty and Scope of the Work}
We propose a GITz model by specifically controlling the RA as a function of frequency of the incident signal, conductivity of graphene and other relevant physical parameters of GIRS, and deriving a closed-form expression of RA. This framework would be beneficial for the GIRS designers to mathematically analyze various problems like maximizing the communication system throughput, proposing an efficient beamforming design, obtaining the optimal number of IRS elements for maximum received power. \textit{To our best knowledge, this is the first work that mathematically relates the GITz design-specific controlling quantity RA, with other graphene material parameters along with the impact of IF}.

\subsection {Major Contributions of the Work}
Key contributions of this work are as follows: 1) A GITz design with material characterization is proposed. 2) Phase dependent IRS controlling parameter RA is modelled in terms of IF and other material parameters like $E_F$, conductivity, charge relaxation time, and GIRS patch geometry. 3) Closed-form expression is derived for RA, essential for mathematically framing a metric to evaluate the performance of GITz system. 4) Extensive numerical results are presented with key insights to analyze the proposed model.

\section{GITz System Description}
Here, we first discuss the characteristics of graphene for enabling an efficient IRS design. Then we describe the GITz communication system considered in this work.

\subsection{Material Characteristics of Graphene}
 Graphene has attracted huge attention in the research community due to its very interesting and unusual electrical properties that could be promising for a wide range of applications in THz band, listed as under:
 \begin{itemize}
     \item It has high carrier mobility making it a very suitable option for flexible electronics \cite{Flexibility}. Moreover, it is thinnest known material with one of the highest electron mobility thereby making undoubtedly the best candidate for THz frequency wireless communication systems.
    
     \item It is highly conductive both thermally as well as electrically, which can be used for designing a tunable resonant element, for example in GIRS system.  
     
     \item Reflection/ absorption at a surface is also subject to the bandgap of the specified material; a photon whose energy is less than the bandgap of material will be reflected. This results in a highly reflective GIRS with sufficient thickness and small bandgap \cite{Joint}.
     
     \item Tunable surface conductivity of 2D graphene sheet can be achieved by shifting of the Fermi energy level or chemical potential from the Dirac point \cite{GRAPHENE} (at the intersection of the conduction band and the valence band).
     
     \item The conductivity of graphene  also plays a dominant role in SPP wave propagation \cite{SPP_THz_Graphene}. Actually, SPP has a key property of electronic excitation due to interaction between surface electric charges at the junction of metal and dielectric substrate in THz frequency range. 
\end{itemize}

\subsection{System Description}
\begin{figure}[h]
    \centering
    \includegraphics[width=3.48in]{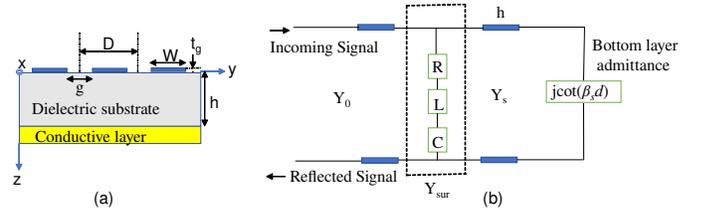}
    \caption{GIRS architecture: (a) Side view of the GIRS model (b) Circuit model for the GIRS design.}
    \label{fig1}
 \end{figure}

GIRS is a tiny, nano-particle sized planar layer which is used to control the EM wave propagation. The dimensional architecture of GIRS is depicted in Fig. \ref{fig1}. The graphene surface is used as an IRS reflector unit cell on top of a dielectric substrate as shown in Fig. \ref{fig1}(a). The desired RA is achieved by varying the biasing voltage applied to the graphene sheet. The unit cell patch dimensions are $W \times W$ arranged with a periodicity $D$. The dielectric substrate is sandwiched between graphene layer and bottom metallic layer with height $h$. The equivalent circuit model of GIRS is illustrated in Fig. \ref{fig1}(b), where top layer of patch creates $R-L-C$ series circuit and $Y_{sur}$ represents the effect of surface admittance of the graphene sheet. Further, an additional dielectric layer provides a parallel circuit to it.

\begin{figure}[h]
    \centering
    \includegraphics[width=3.48in]{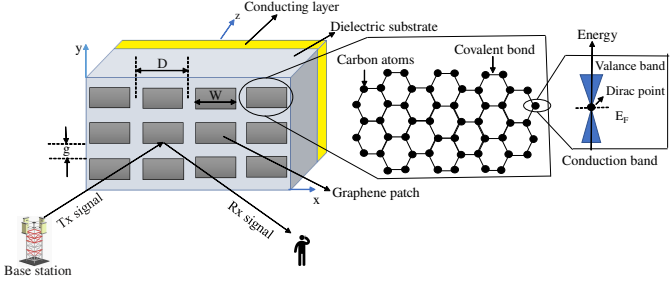}
    \caption{GITz system model.}
    \label{fig2}
 \end{figure}

The physical structure of the graphene metasurface-based reflector design in Fig. \ref{fig2} depicts GIRS-aided communication between a base station as a transmitter (Tx) and a single mobile user as receiver (Rx). In this model, GIRS layer acts as a reflecting surface which can regulate the RA of incoming wave, thereby setting up the controlled Tx-Rx communication pathway. In general, a GIRS is composed of arrays of reflecting elements with a periodic arrangement of graphene-made conductive patches on the upper layer in addition to a metal layer on its bottom. In the same figure, a two dimensional honeycomb lattice nano-structure of graphene is also depicted illustrating the molecular bond structure, made up of carbon atoms having covalent bond. The single carbon atom layout is also depicted, where the conduction and valance bands meet together at the Dirac point. Depending upon the Fermi-level or biasing voltage, charge carriers excite in the material. 
\section{Existing and Proposed GIRS Design Models}
In a IRS-assisted wireless communication system design, the primary aim is to maximize the reflection of wave incident directly on IRS from Tx in a controlled manner. Here, we initially highlight an existing GIRS model, based on PS at Rx side. Next we propose the RA-specific GITz model as a function of various material and electrical parameters.

\subsection{Existing GIRS Model}
At THz frequency, the complex surface conductivity of graphene $sigma_{\text{g}}$ is expressed using Kubo's formula as \cite{Joint},

\begin{equation}\label{cond}
\sigma_{\text{g}} = \frac{2e^2}{\pi \hbar^2}K_{B}T \ln\left [ 2\cosh\left ( \frac{E_{F}}{2K_{B}T} \right )\right ] \frac{j}{\omega +j \tau^{-1}},
\end{equation}

\noindent where $e\triangleq 1.6 \times 10^{-19}$ is the electron charge, $\hbar$ is the reduced Planck's constant, $K_{B}$ is the Boltzmann constant, $T$ is temperature, $E_{F}$ is the Fermi-level, $\omega \triangleq 2\pi f$ is the angular frequency, $j= \sqrt{-1}$, and $\tau$ is the relaxation time. Now, this Fermi-level can be further shifted by varying carrier density of graphene $n_d$ as $\left | E_{F} \right | = \hbar\vartheta_{F}\sqrt{\pi n_{d}}$, where $\vartheta_{F}$ is the Fermi velocity. The carrier density can be modified by adjusting the gate bias voltage $V_g$, i.e. $n_{d}= \sqrt{n_{0}^2+\alpha \left | V_{CNP}-V_{\text{g}} \right |^2}$, where $n_{0}$ indicates the residual carrier density in puddle formation of electron and hole near the Dirac point in the graphene material, $\alpha$ is the gate capacitance in the electrode and $V_{CNP}$ is the compensating voltage \cite{Joint}. So, $\sigma_g$ can be varied and tuned in accordance with the applied voltages \cite{Gate_volt}, thereby forming the basis of designing an electrically controlled GIRS.

In GIRS, at the boundary between a metal and dielectric, the coupling of external EM waves and plasmons on the metal surface generates a wave that propagates along the interface \cite{SPP}. The EM responses of the IRS reflecting elements govern the resonance of GIRS caused by the constructive or destructive interference of multiple reflections. This results into a PS $\theta$ at the Rx side of GIRS-assisted communication system \cite{joint23},
\begin{equation} \label{theta}
    \theta = m\pi-Wk_0\text{Re}\left(\sqrt{1-\left(\frac{2}{n_0\sigma_{\text{g}}}\right)^2}\right)\triangleq \theta(E_F, f, \tau, n_0),
\end{equation}
where $\text{Re}(\cdot)$ represents the real part of a complex number, $W$ is the width of graphene patch, $m$ is an integer, $k_{0} \triangleq 2\pi/ \lambda$ is the free space wave number, $\lambda$ is the wavelength, and $n_0$ is free space impedance (considered as $377$ $\Omega$ in this work). Hence, from \eqref{cond} and \eqref{theta}, PS associated with a GIRS can be altered by varying the conductivity of graphene and it is a function of ($E_F, f, \tau, n_0$).

\subsection{Proposed RA-based GITz Characterization}
Here we propose a GITz design model based on the phase-dependent RA. Specifically, we aim to obtain a closed-form expression of RA of the signal reflected by GIRS elements in terms of various material properties and incident signal frequency. This relationship would help in controlling IRS by selecting the required PS and RA. Motivated by the fact that effective admittance of an IRS element and other physical properties of graphene material can control the signal propagation, we start with the fundamental definition of reflection coefficient $\Gamma$ corresponding to a reflecting surface \cite{cai},
\begin{equation}\label{gamma1}
    \Gamma= \mathcal{R} e^{j\theta},
\end{equation}
where $\mathcal{R}\in [0,1]$ is the RA and $\theta \in [-\pi, \pi]$ is the PS. Moreover, based on the equivalent circuit model of a GIRS (c.f. Fig. 1(b)), the underlying $\Gamma$ can be expressed as \cite{Ysur}, 
\begin{equation}\label{gamma2}
     \Gamma = \left(\frac{Y_{0}-Y_{\text{in}}}{Y_{0}+Y_{\text{in}}}\right),
 \end{equation}
 where $Y_{0}$ is the characteristic admittance and $Y_{in}$ is the input admittance of the circuit that is calculated as,
 \begin{equation}\label{yin}
     Y_{\text{in}} = \mathcal{Y}- jY_{\text{s}}\cot(\beta_{s}d),
 \end{equation}
 where  $Y_{s} \triangleq n_{s}/n_0$ is the ratio of dielectric slab refractive index $n_{s}$ and free-space refractive index $\eta_0$, $\beta_{s}\triangleq k_{o}n_{s}$ is the propagation constant, $d= \lambda/4$ is the quarter wavelength. Finally, $\mathcal{Y}$ is the surface admittance of graphene patch element which depends upon the material geometry and its conductivity, defined as \cite{Ysur},
  \begin{equation} \label{Ysur}
     \mathcal{Y} = \frac{\mathcal{S}^2}{D^2 \mathcal{K}}\left(\sigma_{\text{g}}^{-1}+\frac{q}{j\omega \epsilon_{\text{eff}}}\right)\triangleq \mathcal{Y}(\sigma_g, t_g),
 \end{equation}
 where $\mathcal{S}=1.015$ $W^2$, $\mathcal{K}=1.163$ $W^2$, $q$ is the eigen value (varies between $0.186$ $\pi$/$W$and $0.256$ $\pi$/$W$), $W$ is the patch-width, $\epsilon_{\text{eff}}$ is the effective permittivity of graphene governed by the relation, $D$ is periodicity of the patch, $\epsilon_{\text{eff}} \triangleq 1+ \left(\frac{j\sigma_{\text{g}}}{\omega \epsilon_{0} t_{g}}\right)$ with $t_{g}$ being the thickness of the graphene patch.
 
 With the above analysis, RA of the signal received at the destination user in a GITz system can be obtained from \eqref{gamma1}, \eqref{gamma2} and \eqref{yin} as,

\begin{equation} \label{RA2}
    \mathcal{R}\hspace{-0.5mm} =\hspace{-0.5mm} \left[\frac{Y_0\hspace{-0.5mm}-\hspace{-0.5mm} \{\mathcal{Y}\hspace{-0.5mm}-\hspace{-0.5mm}jY_{s} \cot{(\beta_{s} d)\} }}{Y_0\hspace{-0.5mm}+ \hspace{-0.5mm}\{\mathcal{Y}\hspace{-0.5mm}-\hspace{-0.5mm}jY_{s} \cot{(\beta_{s} d)\}}} \right]\hspace{-0.5mm} e^{-j\Theta}\triangleq \mathcal{R}(f,E_F, \tau, n_0, t_g).
\end{equation}
 
 Hence, a closed-form expression of RA $\mathcal{R}(f,E_F, \tau, n_0, t_g)$ has been achieved by \eqref{RA2} for a GIRS design meant for efficient wireless communication in THz frequency band.
 
\section{Results and Discussion}
In this Section, we provide numerical results for design and analysis of the proposed RA-based GITz model. We consider a GIRS patch with period of $90$ $\mu$m. For simulations, unless specified exclusively, we have used system parameter values as: $f = \{1, 2, 3\}$ THz, $n_0 = 377$ $\Omega$, charge relaxation time is assumed to be $\tau=6$ ps, $E_{F} \in[0.5,2.5]$ eV, $ W\in [10,25]$ $\mu m$, $T=300$ K and phase $\theta$ changes from $-\pi$ to $\pi$.  
 \begin{figure}
    \centering
    \includegraphics[width=3.48in]{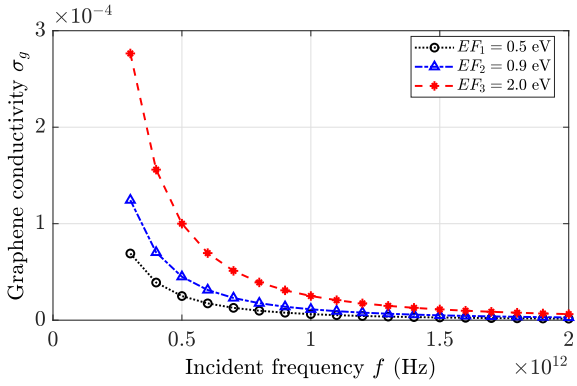}
    \caption{Effect of IF on graphene conductivity.}
    \label{fig8}
\end{figure}
We begin with Fig. \ref{fig8} by first validating the most fundamental relationship of material conductivity with incident signal frequency, Fermi-level and other material parameters, provided by \eqref{cond}.  Performance of GITz model is analyzed using the conductivity of graphene against different IF in Fig. \ref{fig8}. Clearly, conductivity tends to decrease with IF, but increase with increasing Fermi-levels, as also evident from the proposals of other existing works \cite{Ysur, Joint, Hybrid}. In this work, we have used MATLAB software to simulate the mathematical expression obtained in Section III.

\begin{figure}
    \centering
    \includegraphics[width=3.48in]{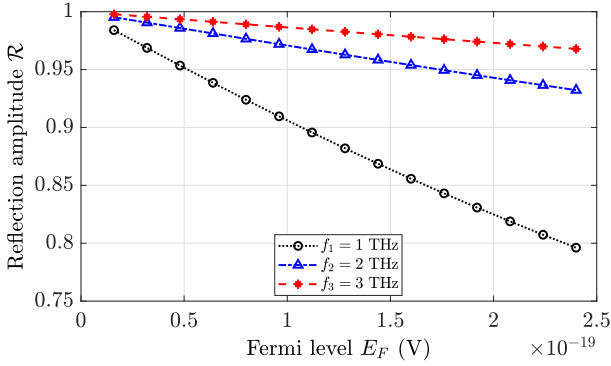}
    \caption{Effect of different Fermi-levels on  RA.}
    \label{fig3}
\end{figure}

The next set of results correspond to further dependence of $\sigma_g$ on various other relevant material parameters like $E_F, W, \tau, n_0$, whose effect on RA of a GITz system has not been explored in the current literature. Fig. \ref{fig3} shows the impact of $E_F$ on RA of GIRS elements at different IFs. All the curves are observed to be inline with \eqref{RA2}. The highest value of $\mathcal{R}$ has been found to be $0.99$ at $E_{F}=0.16$ eV with $f=3$ THz. 

\begin{table*}[b]
\centering
\caption{Comparative analysis of different conducting materials against graphene}
{
\begin{tabular}{|l|l|l|l|l|l|} 
\hline
Conducting material & \begin{tabular}[c]{@{}l@{}}Conductivity\\(MS/m)\end{tabular} & \begin{tabular}[c]{@{}l@{}}Thickness\\($\mu$m)\end{tabular} & \begin{tabular}[c]{@{}l@{}}Electron mobility\\{(cm$^2$/Vs)}\end{tabular} & Melting point (K) & Price          \\ 
\hline
Cu                  & 58                                                           & 35                                                                        & \multirow{3}{*}{30-50}                                                 & 1357              & Low            \\ 
\cline{1-3}\cline{5-6}
Au                  & 45                                                           & 0.1 \cite{Au_Thickness}                                                                      &                                                                        & 1337     & High           \\ 
\cline{1-3}\cline{5-6}
Ag                  & 61                                                           & 9 \cite{Ag_Thickness}                                                                        &                                                                        & 1253              & Moderate high  \\ 
\hline
Graphene \cite{Graphenea, conductivity, M_Temp}            & up to 100                                                    & 0.000345                                                                  & 10000                                                                  & 3000-7000         & Very high      \\
\hline
\end{tabular}
}
\end{table*}

\begin{figure}
    \centering
    \includegraphics[width=3.48in]{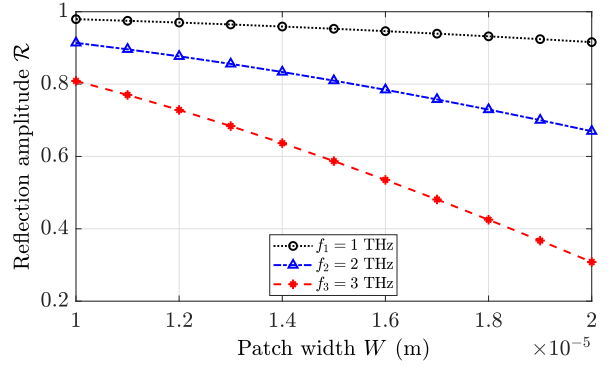}
    \caption{Effect of $W$ on RA for various IF.}
    \label{fig4}
\end{figure}

\begin{figure}
    \centering
    \includegraphics[width=3.48in]{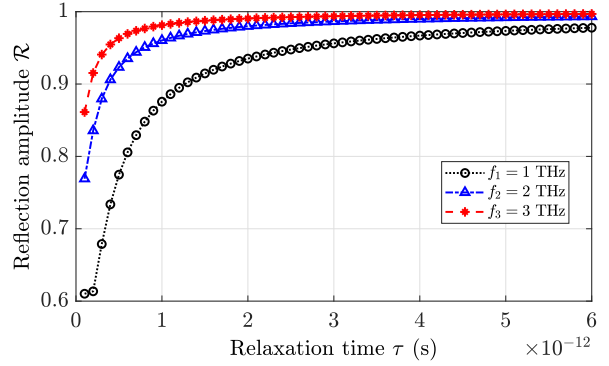}
    \caption{Impact of different relaxation time on RA.}
    \label{fig5}
\end{figure}

The effect of graphene patch width on RA of GIRS element is presented in Fig. \ref{fig4} for different sets of IF. Here, the value of $E_{F}$ is fixed at $1.25$ eV. It is clear that if the $W$ is $10$ $\mu$m then, maximum RA is achieved at the Rx. Moreover, we obtain minimum amplitude of $0.3$ with $W=20$ $\mu$m at the Rx end. Note that decreasing the patch width leads to a better amplitude of reflection at low IF. Likewise, Fig. \ref{fig5} depicts the impact of varying the charge relaxation time of graphene on GIRS RA. We observe that as increase in the relaxation time is responsible for enhancement in the RA value. The highest amplitude is observed  at $\tau=6$ ps at higher IF $f=3$ THz.

\begin{figure}
    \centering
    \includegraphics[width=3.48in]{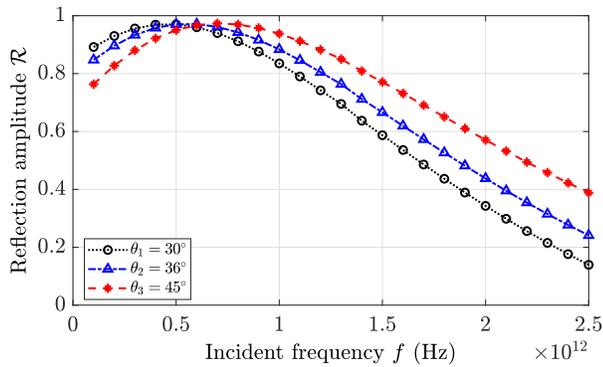}
    \caption{Influence of IF on RA for different PS.}
    \label{fig6}
\end{figure}

 \begin{figure}
    \centering
    \includegraphics[width=3.48in]{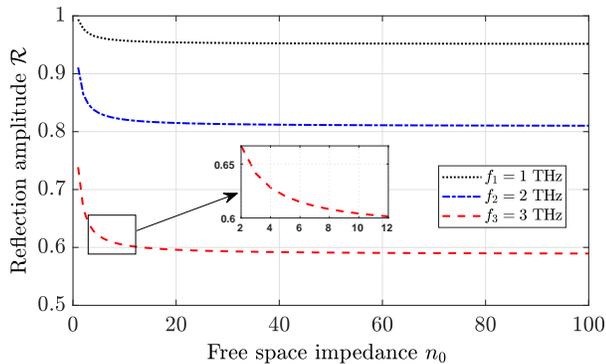}
    \caption{Impact of free space impedance on RA.}
    \label{fig7}
\end{figure}
Next in Fig. \ref{fig6}, for a GIRS designer, we illustrate the effect of varying phase response and IF on RA at the receiver of GITz system. In accordance with \eqref{RA2}, we observe that initially $\mathcal{R}$ increases with $f$ till an optimum value i.e. $f=0.6$ THz, after which it starts decreasing. Interestingly, beyond $f=0.6$ THz, the behavior of RA is also conversed with respect to the induced PS, suggesting thereby RA can be improved at low PS values if the incident signal frequency is also low.

In Fig. \ref{fig7} we plot the impact of $n_0$ of GIRS patch on RA for various IFs, with fixed value of Fermi-level $E_{F}=2.5$ eV. The effect of medium on RA is clearly visible in low $n_{0}$ region. Here, RA is reduced with increasing $n_{0}$ but soon becomes constant at high $n_{0}$. So, the influence of free space impedance is not very significant on the efficiency of reflection of a GIRS, if its value is taken sufficiently high. 

Finally in Table I, different conducting materials and their comparative performance analysis are presented against graphene. It is clear that although graphene does not appear to be a cost-effective material as compared to other conventional materials loaded with parasitic elements, graphene-made patch has superb electron carrying capacity, larger conductivity than other extensively available materials.

\section{Conclusion}
We proposed a GIRS material characterization model for efficient THz communications. We established a mathematical relationship of RA with IF and various electrical and material-specific physical parameters corresponding to a GITz system. Further, we provided simulation results and observed the effect of Fermi-level, patch geometry and charge relaxation time on RA at different THz frequencies. This work will serve as a benchmark for evaluating the performance of IRS-based high frequency communications. In future, we intend to utilize the proposed GITz model for maximizing the throughput of communication system with optimal PS and RA.

% , thereby dealing with the associated optimization problems of maximum capacity, optimal GIRS placement, effective beamforming design for GIRS and many more.
%%%%%
\makeatletter
\renewenvironment{thebibliography}[1]{%
  \@xp\section\@xp*\@xp{\refname}%
  \normalfont\footnotesize\labelsep .5em\relax
  \renewcommand\theenumiv{\arabic{enumiv}}\let\p@enumiv\@empty
  \vspace*{-1pt}% NEW
  \list{\@biblabel{\theenumiv}}{\settowidth\labelwidth{\@biblabel{#1}}%
    \leftmargin\labelwidth \advance\leftmargin\labelsep
    \usecounter{enumiv}}%
  \sloppy \clubpenalty\@M \widowpenalty\clubpenalty
  \sfcode`\.=\@m
}{%
  \def\@noitemerr{\@latex@warning{Empty `thebibliography' environment}}%
  \endlist
}
\makeatother

\bibliographystyle{IEEEtran}
\bibliography{reference}
\end{document}